\documentclass[twocolumn]{aastex63}

\newcommand\teff{T_{\text{eff}} }

\newcommand{\kms}{km\,s$^{-1}$}

\usepackage{textcomp}

\shorttitle{LP\,398-9: Circumstellar Material and Rotation}
\shortauthors{Chandra et al.}

\graphicspath{{./}{fig/}}

\begin{document}
 
\title{The SN\,Ia Runaway LP\,398-9: Detection of Circumstellar Material and Surface Rotation}

\correspondingauthor{Vedant Chandra}
\email{vedant.chandra@cfa.harvard.edu}

\author[0000-0002-0572-8012]{Vedant~Chandra}
\affiliation{Center for Astrophysics $\mid$ Harvard \& Smithsonian, 60 Garden St, Cambridge, MA 02138, USA}
\affiliation{Department of Physics \& Astronomy, Johns Hopkins University, 3400 N Charles St, Baltimore, MD 21218, USA}

\author[0000-0003-4250-4437]{Hsiang-Chih~Hwang}
\affiliation{Institute for Advanced Study, Einstein Drive, Princeton, NJ 08540, USA}
\affiliation{Department of Physics \& Astronomy, Johns Hopkins University, 3400 N Charles St, Baltimore, MD 21218, USA}

\author[0000-0001-6100-6869]{Nadia~L.~Zakamska}
\affiliation{Department of Physics \& Astronomy, Johns Hopkins University, 3400 N Charles St, Baltimore, MD 21218, USA}
\affiliation{Institute for Advanced Study, Einstein Drive, Princeton, NJ 08540, USA}

\author[0000-0002-9632-1436]{Simon~Blouin}
\affiliation{Los Alamos National Laboratory, PO Box 1663, Los Alamos, NM 87545, USA}
\affiliation{Department of Physics and Astronomy, University of Victoria, Victoria, BC V8W 2Y2, Canada}

\author[0000-0001-6515-9854]{Andrew~Swan}
\affiliation{Department of Physics \& Astronomy, University College London, Gower Street, London WC1E 6BT, UK}

\author[0000-0003-1394-4644]{Thomas~R.~Marsh}
\affiliation{Department of Physics, University of Warwick, Coventry, CV4 7AL, UK}

\author[0000-0002-9632-6106]{Ken~J.~Shen}
\affiliation{Department of Astronomy and Theoretical Astrophysics Center, University of California, Berkeley, CA 94720, USA}

\author[0000-0002-2761-3005]{Boris~T.~G{\"a}nsicke}
\affiliation{Department of Physics, University of Warwick, Coventry, CV4 7AL, UK}

\author[0000-0001-5941-2286]{J.~J.~Hermes}
\affiliation{Department of Astronomy \& Institute for Astrophysical Research, Boston University, 725 Commonwealth Ave., Boston, MA 02215, USA}

\author[0000-0002-8935-0431]{Odelia~Putterman}
\affiliation{Department of Astronomy \& Institute for Astrophysical Research, Boston University, 725 Commonwealth Ave., Boston, MA 02215, USA}

\author[0000-0002-4791-6724]{Evan~B.~Bauer}
\affiliation{Center for Astrophysics $\mid$ Harvard \& Smithsonian, 60 Garden St, Cambridge, MA 02138, USA}

\author[0000-0003-2427-4287]{Evan~Petrosky}
\affiliation{Department of Physics \& Astronomy, Johns Hopkins University, 3400 N Charles St, Baltimore, MD 21218, USA}
\affiliation{Department of Physics, University of Michigan, 450 Church St, Ann Arbor, MI 48109, USA}

\author[0000-0001-5944-4480]{Vikram~S.~Dhillon}
\affiliation{Department of Physics and Astronomy, University of Sheffield, Sheffield S3 7RH, UK}
\affiliation{Instituto de Astrof\'{i}sica de Canarias, E-38205 La Laguna, Tenerife, Spain}

\author[0000-0001-7221-855X]{Stuart~P.~Littlefair}
\affiliation{Department of Physics and Astronomy, University of Sheffield, Sheffield S3 7RH, UK}

\author[0000-0002-1210-4144]{Richard~P.~Ashley}
\affiliation{Department of Physics, University of Warwick, Coventry, CV4 7AL, UK}
\affiliation{Isaac Newton Group of Telescopes, Apartado de Correos 321, Santa Cruz de La Palma, E-38700, Spain}

\begin{abstract}

\noindent A promising progenitor scenario for Type Ia supernovae (SNeIa) is the thermonuclear detonation of a white dwarf in a close binary system with another white dwarf. After the primary star explodes, the surviving donor can be spontaneously released as a hypervelocity runaway. One such runaway donor candidate is LP\,398-9, whose orbital trajectory traces back $\approx 10^5$ years to a known supernova remnant. Here we report the discovery of carbon-rich circumstellar material around LP\,398-9, revealed by a strong infrared excess and analyzed with follow-up spectroscopy. The circumstellar material is most plausibly composed of inflated layers from the star itself, mechanically and radioactively heated by the past companion's supernova. We also detect a $15.4$\,hr periodic signal in the UV and optical light curves of LP\,398-9, which we interpret as surface rotation. The rotation rate is consistent with theoretical predictions from this supernova mechanism, and the brightness variations could originate from surface inhomogeneity deposited by the supernova itself. Our observations strengthen the case for this double-degenerate SNIa progenitor channel, and motivate the search for more runaway SNIa donors. 

\end{abstract}

\keywords{white dwarfs, supernovae: general, circumstellar matter, stars: rotation}

\section{Introduction}\label{sec:intro}

Type Ia supernovae (SNe\,Ia) are luminous transients that are valuable standard candles to measure cosmological parameters \citep[e.g.,][]{Riess1998, Perlmutter1999, Riess2019}, and play a crucial role in the chemical evolution of stellar populations \citep[e.g.,][]{Tinsley1979, Matteucci2009, Kirby2019}. Despite the importance of SNeIa, their progenitor scenario is uncertain, and is the subject of intense theoretical and observational efforts \citep[see e.g.,][for a review]{Hillebrandt2000, Maoz2013,Soker2019}. It is generally accepted that SNe\,Ia originate from massive white dwarfs (WDs) that accrete matter from binary companions, but the nature and fate of the donor companion is unknown. 

For several decades, the leading hypothesis was the `single degenerate' scenario \citep[e.g.,][]{Whelan1973}, in which a carbon--oxygen core (C/O) WD accretes matter from a nondegenerate star until it approaches the Chandrasekhar limit \citep{Chandrasekhar1931}. However, it is challenging for the accreting WD to gain mass in the first place \citep[e.g.,][]{Nomoto1982, Iben1984}, and there is a growing body of observational evidence that suggests most SNe\,Ia cannot originate from single-degenerate progenitors \citep[e.g.,][]{Kerzendorf2009, Kasen2010, Li2011, Bloom2012, Margutti2012, Woods2017}. An alternative is the `double degenerate' scenario in which two WDs could either merge and consequently explode \citep[e.g.,][]{Iben1984, Webbink1984}, or have one WD detonate after mass transfer \citep[e.g.,][]{Bildsten2007}. The explosion mechanism can be a double-detonation, during which a detonation in the outer helium shell induces a second detonation in the carbon core that is powerful enough to unbind the star \citep{Taam1980, Guillochon2010, Dan2011, Raskin2012, Pakmor2013, Shen2014}. 

In the double-degenerate case, a double-detonation can occur during the mass transfer phase, well before the two WDs merge. In this `dynamically driven double-degenerate double-detonation' (D$^6$) scenario, the primary WD can explode well below the Chandrasekhar limit  \citep{Pakmor2013, Shen2018a, Shen2018, Tanikawa2018, Tanikawa2019}. Observational studies have independently found that a significant fraction of SNeIa could originate from sub-Chandrasekhar mass WDs \citep[][]{Scalzo2014, Dhawan2017, Kirby2019, DelosReyes2020, Sanders2021}. Recent theoretical work has shown that the D$^6$ mechanism can reproduce the observational signatures of most SNIa \citep{Shen2021a,Shen2021c}. In this scenario, the donor WD could survive the explosion of the primary if it occurs in the early stages of mass transfer \citep{Shen2017a}. The binary orbit becomes spontaneously unbound after the supernova, ejecting the donor WD at the orbital velocity, $\gtrsim 1000\ \text{km s}^{-1}$. 

\cite{Shen2018} discovered three candidates that appear to be `runaway' WD donors ejected from SNe\,Ia that exploded via the D$^6$ mechanism. These stars are among the fastest unbound stars in the Galaxy, with estimated space velocities $\gtrsim 1000$\,\kms{}. Their radii are inflated by an order of magnitude compared to typical WDs, likely due to deposited energy from the SN\,Ia ejecta. These stars do not fit into any classical spectroscopic WD subtypes, due to their much lower apparent surface gravity and peculiar composition. All three candidates in \cite{Shen2018} have similar low-resolution optical spectra, with absorption signatures of carbon, oxygen, magnesium, and calcium. One of these candidates is LP\,398-9, referred to as `D6-2' in \cite{Shen2018}. Remarkably, LP\,398-9's inferred orbital trajectory extrapolated back $9 \times 10^4$ years passes through the on-sky position of a known supernova remnant G70.0-21 \citep{Fesen2015,Shen2018,Raymond2020}. G70.0-21 lies far off the Galactic plane, suggesting that it is a Type Ia remnant, and its shock velocity-inferred distance is $1-2$~kpc, consistent with LP\,398-9's distance of $\approx 1$~kpc.

Here we present follow-up observations of LP\,398-9 that reveal the presence of significant quantities of circumstellar material, as well as a 15.4\,hr photometric period. We argue that both of these observables can be linked to the D$^6$ origin of the system, strengthening the case for this SNIa progenitor channel. After summarizing our data and observations in $\S$\ref{sec:data}, we present our analysis in $\S$\ref{sec:analysis}. We present our results in $\S$\ref{sec:results}, and discuss our findings in $\S$\ref{sec:discussion}. 

\section{Data}\label{sec:data}

In this section we describe the archival data we collected for LP\,398-9, as well as our own follow-up observations. In $\S$\ref{sec:data.phot} we assemble the spectral energy distribution (SED) from archival data. In $\S$\ref{sec:data.lc} we describe our light curves and follow-up photometry, and in $\S$\ref{sec:data.spec} we describe our follow-up spectroscopy.

\subsection{Spectral Energy Distribution}\label{sec:data.phot}

\begin{figure}
    \centering
    \includegraphics[width=\columnwidth]{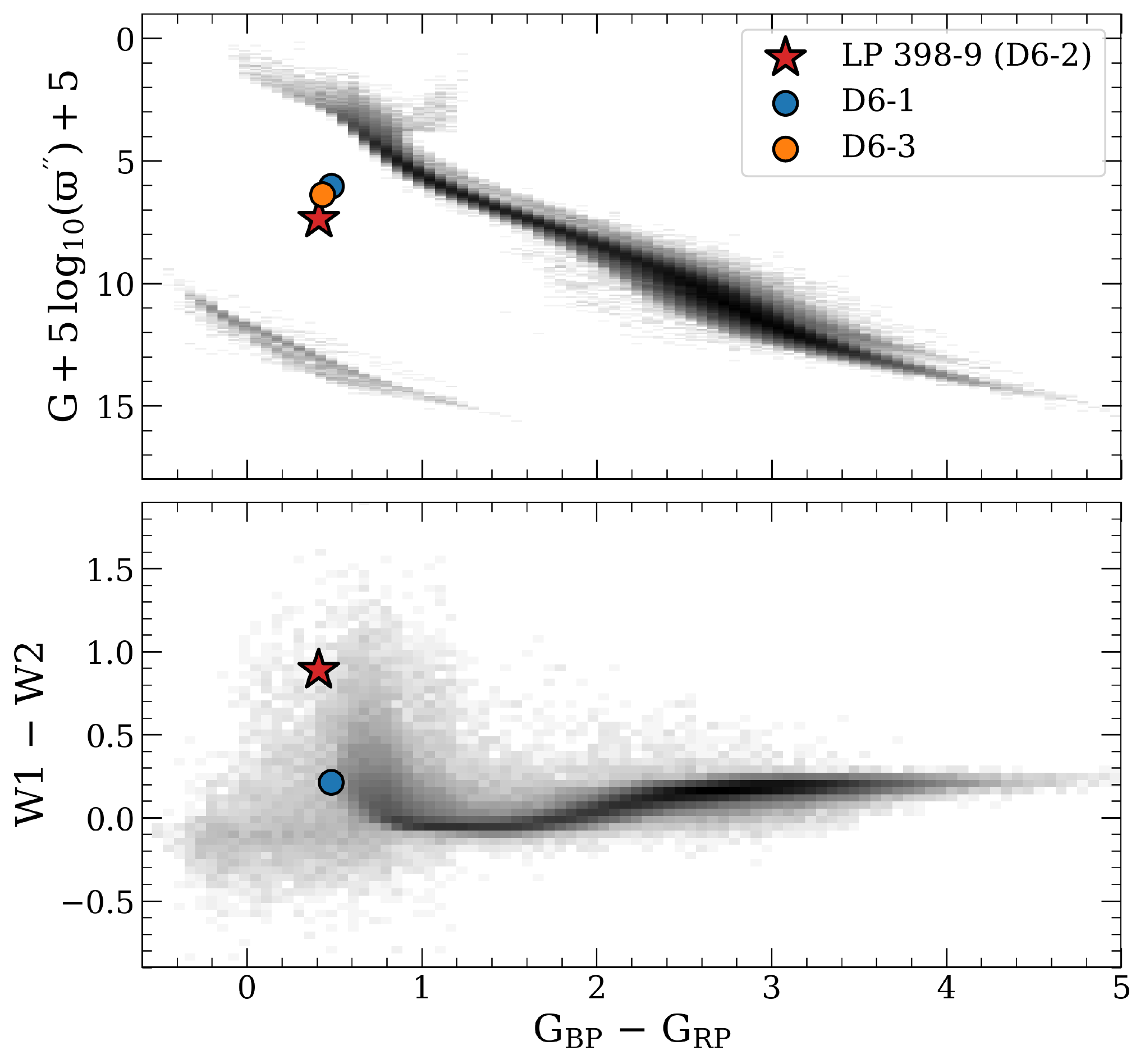}
    \caption{Top: Location of LP\,398-9 on the \textit{Gaia} EDR3 color-magnitude diagram. The background sample consists of stars within 100pc of the Sun \citep{GaiaCollaboration2020}. Bottom: The \textit{Gaia}-\textit{WISE} color-color space, with the same background sample of nearby stars. LP\,398-9 has a significant excess in the \textit{W1}-\textit{W2} color compared to other stars of a similar optical color. For comparison, we show the other two D$^6$ candidates from \cite{Shen2018}. `D6-3' does not have secure WISE photometry due to a crowded field, and is consequently absent from the bottom panel.}
    \label{fig:cmd}
\end{figure}

LP\,398-9 has reliable archival photometry in the \textit{GALEX} \textit{NUV} \citep{Martin2005, Million2016}, Sloan \textit{ugriz} \citep{Fukugita1996, Gunn1998, Doi2010, Blanton2017, Ahumada2020}, 2MASS \textit{JH} \citep{Skrutskie2006}, and \textit{WISE} \textit{W1,W2} \citep{Wright2010, Mainzer2011, Mainzer2014} bands. The detections in \textit{GALEX} \textit{FUV}, 2MASS \textit{Ks}, and \textit{WISE} \textit{W3,W4} are unreliable or absent. LP\,398-9 also has secure astrometry from the \textit{Gaia} space observatory Early Data Release 3 (EDR3;  \citealt{Gaia2016, Gaia2018, GaiaCollaboration2021, Lindegren2021}), with $\varpi/\sigma_\varpi \sim 18$. We correct the observed photometry for interstellar extinction along the line of sight using 3D dust maps \citep{Green2018dm, Green2015, Green2018, Green2018b} queried at the \textit{Gaia} inverse-parallax distance of 840 pc. We adopt the extinction law of \cite{Fitzpatrick2007} with $R_{\text{V}}=3.1$. The \textit{Gaia} color--magnitude diagram and \textit{Gaia}--\textit{WISE} color--color diagram of LP\,398-9 are shown in Figure \ref{fig:cmd}. The lower panel shows LP\,398-9's infrared excess compared to other stars at a similar optical color. Of the other two D$^6$ candidates from \cite{Shen2018}, D6-1 has no $\mathit{W1}-\mathit{W2}$ color excess, and D6-3 is in a crowded field and consequently does not have secure \textit{WISE} photometry. 

\begin{figure}
    \centering
    \includegraphics[width=\columnwidth]{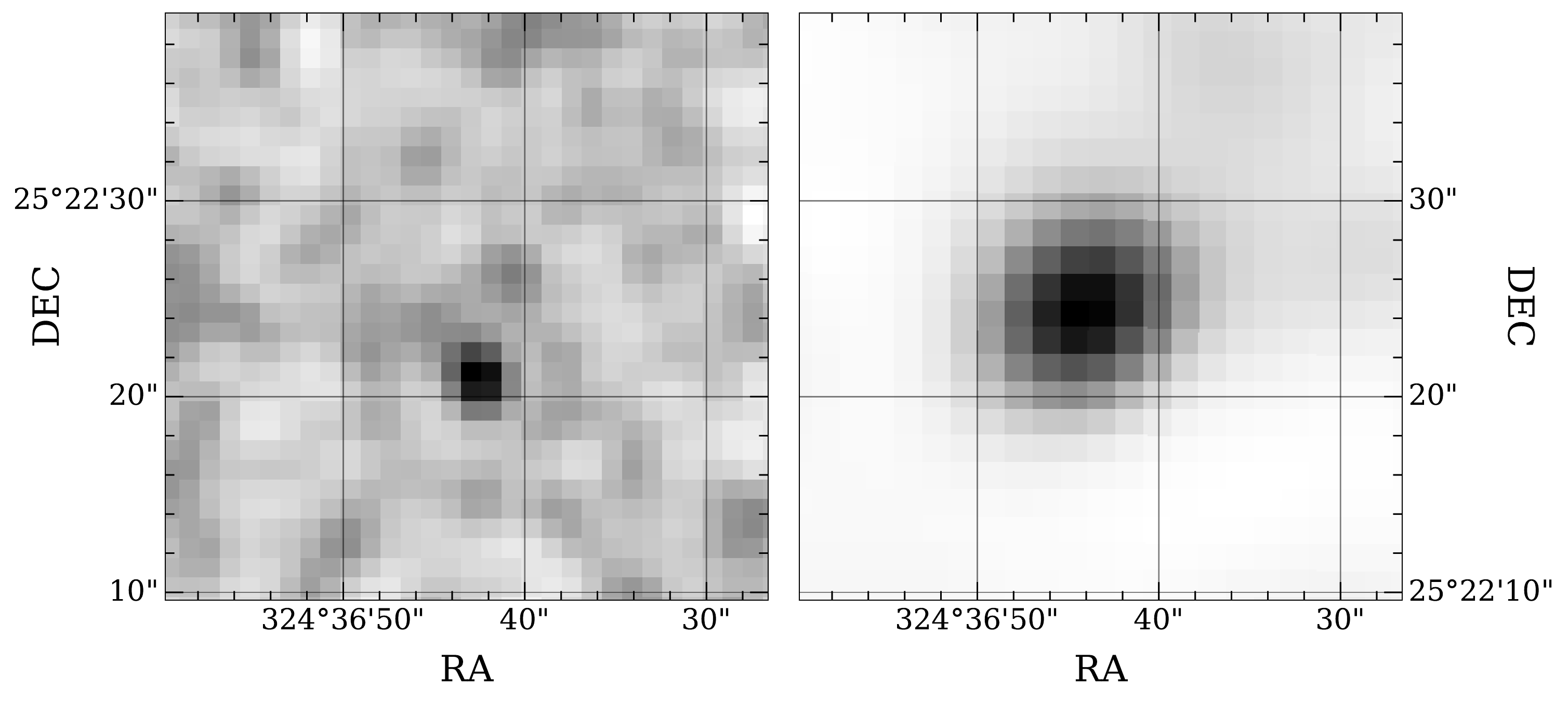}
    \includegraphics[width=\columnwidth]{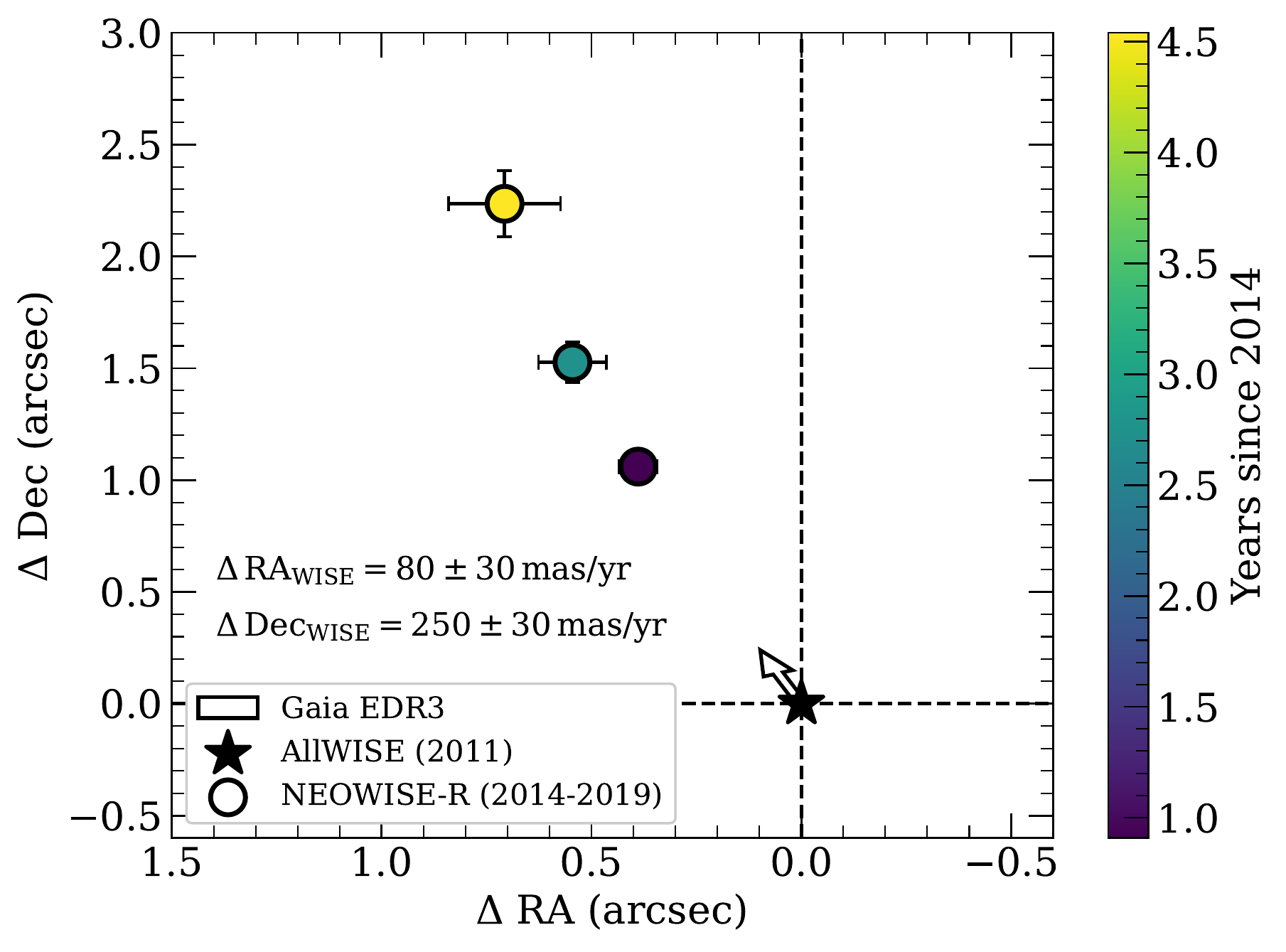}
    \caption{Top left: 2MASS $J$ band image from $\approx 2001$. Top right: AllWISE channel \textit{W1} image ten years later, from $\approx 2011$. The centroid has clearly shifted by several arcseconds towards the NE direction. Bottom: Position of LP\,398-9 on the sky during the course of 5 years of the NEOWISE mission. Each datapoint is the average of 20 individual measurements. The arrow indicates 1 year of the \textit{Gaia} EDR3 proper motion, centred on the AllWISE coordinate. There is no bright contaminant within 6 arcseconds and the \textit{WISE} proper motion is consistent with the optical proper motion from \textit{Gaia}, supporting the fact that the IR excess and variability comes from LP\,398-9 itself.}
    \label{fig:source_conf}
\end{figure}

One potential contaminant of \textit{WISE} imaging is source confusion due to the relatively coarse angular resolution of \textit{WISE} \citep[see e.g.,][]{Dennihy2020a}. In Figure \ref{fig:source_conf} we compare the \textit{J}-band image from 2MASS to the AllWISE image in channel \textit{W1}. In the 2MASS image, there is no discernible background source within 6 arcseconds of LP\,398-9. Furthermore, because the \textit{WISE} data of LP\,398-9 appears time-variable ($\S$\ref{sec:analysis.rot}), we exclude the possibility of a background blazar by finding no radio counterpart in the NRAO VLA Sky Survey (NVSS; \citealt{Condon1998}). A conclusive method to verify the association between the \textit{WISE} data and LP\,398-9 is by comparing optical astrometry to the source position measured by \textit{WISE}. In Figure \ref{fig:source_conf} we display time-averaged measurements of the position of LP\,398-9 on the sky as measured over 5 years of the NEOWISE mission (2014-2019), along with the reference position from AllWISE in 2011 \citep{Cutri2012}. We perform a trimmed least-squares linear regression \citep{LTS, Cappellari2013a} on the RA and Dec as a function of time to estimate the \textit{WISE} proper motion. The parallax motion of $\sim 1$ mas (known from \textit{Gaia}) is negligible here. We obtain $\Delta{{\text{RA}_{\text{WISE}}}} = 80 \pm 30$ mas/yr and $\Delta{{\text{Dec}_{\text{WISE}}}} = 250 \pm 30$ mas/yr. This is consistent within 1-$\sigma$ with the \textit{Gaia} EDR3 proper motion of LP\,398-9 (Table \ref{tab:params}). To ensure our method is robust, we compute the \textit{WISE} proper motions of some nearby objects in the field and verify that they too match their \textit{Gaia} data within uncertainties. This astrometric test confirms that the \textit{WISE} photometric source is securely associated with LP\,398-9.

\subsection{Light Curves}\label{sec:data.lc}

In addition to the time-averaged catalog measurements from AllWISE, we utilize individual exposures taken over the past nine years. We assemble the \textit{WISE} light curve of LP\,398-9 by querying the AllWISE \citep{Wright2010} and NEOWISE \citep{Mainzer2011,Mainzer2014} databases for single-exposure photometric measurements. There are 27 AllWISE datapoints from 2010~May~27 to 2010~November~10, and 178 NEOWISE datapoints from 2014~May~31 to 2019~November~10. We perform quality cuts to remove spurious detections with outlying RA/Dec offsets and undefined magnitude uncertainties. The resulting light curve has 168 datapoints over a 9 year baseline. The cadence of the observations is irregular and follows the \textit{WISE} position-dependent scanning strategy, with a $\simeq$ 180 day gap between successive observing runs. Each run is $\approx 1.5$ days long and consists of exposures taken $\approx 1.5$ hours apart. 

We queried archival photometry for LP\,398-9 from the Zwicky Transient Facility (ZTF; \citealt{Bellm2019,Masci2019}). We assembled a light curve consisting of datapoints collected between 2020~January~13 and 2021~April~27 in the $g$ and $r$ photometric bands. We only used the most recent year of ZTF data to minimize the effect of long term brightness variations (discussed further in $\S$\ref{sec:analysis.rot}). To maximize our time coverage, we combined the $g$ and $r$ data after subtracting their respective median magnitudes \citep[e.g.,][]{Burdge2020}. We removed bad datapoints with ZTF quality flag \texttt{catflags > 0} and sharpness \texttt{sharp > 0.5} \citep[e.g.,][]{Guidry2021}. Our final ZTF \textit{gr} lightcurve has 205 datapoints. 

We observed LP\,398-9 with HiPERCAM \citep{Dhillon2016,Dhillon2018,Dhillon2021} on the 10.4-m Gran Telescopio Canarias (GTC) at the Observatorio del Roque de los Muchachos on the island of La Palma, Spain, on 2018~June~11 from 03:38 to 05:15 UTC (PI: Dhillon). We utilized a customised high-throughput approximation of the Sloan $ugriz$ filter set to obtain simultaneous 5-band photometry. A cadence of $1.222$~sec was used for the $g$, $r$ and $i$ arms, while every other readout was skipped on the $u$ and $z$ arms giving a $2.444$~sec cadence. The weather was clear with seeing $\sim$ 0.8\arcsec--1.5\arcsec. The Moon was 80\% illuminated. We debiased and flat-fielded the data using twilight sky flats. We performed comparative aperture photometry using apertures that tracked the target positions with radii set to 1.8 times the mean full width at half-maximum of the stellar profiles. We defer to \cite{Dhillon2021} for further details of the reduction procedures. We used \textit{Gaia} EDR3 1798008691771518464 ($G=15.1$) as the main comparison star, and \textit{Gaia} EDR3 1798020408442309248 ($G=16.5$) to confirm that the main comparison star was not itself variable. We also observed LP\,398-9 with the Wide Field Camera on the 2.5-m Isaac Newton Telescope (INT), on 2018~August~2 from 00:17 to 04:44 UTC. We utilized a Sloan $g$ filter with exposure times of $100$~sec. We performed aperture photometry relative to a comparison star, and defer to \cite{Dhillon2021} for further details on the reduction procedures. 

We observed LP\,398-9 with the IO:O imager on the 2-m Liverpool Telescope (LT; \citealt{Steele2004}) at the Observatorio del Roque de los Muchachos, over the course of a month from 2021~July~8 to 2021~August~8 (PI: G{\"a}nsicke). We obtained a total of 154 images in the Bessel \textit{B} band (effective wavelength $\simeq 4450$\,\AA{}) with an exposure time of 60\,s, averaging approximately five images per night. We performed comparative aperture photometry against the nearby star \textit{Gaia} EDR3 1798008721834715648 ($G=13.8$). 

We also analyze observations of LP\,398-9 obtained with the Space Telescope Imaging Spectrograph (STIS; \citealt{Woodgate1998}) of the \textit{Hubble Space Telescope}(\textit{HST}) over six consecutive orbits on 2020~October~3 (PI: Shen). The spectra are flux-calibrated and cover a wavelength range of 1600--3600\,\AA{} with the G230L grating. We use these spectra to derive an ultraviolet light curve that is analyzed in this paper. A full analysis of the STIS spectrum will be presented in a forthcoming publication. We took advantage of the photon-counting \texttt{TIME-TAG} mode to re-bin the spectra into 9-minute exposure times, and computed the average flux in the 1700--3100\,\AA{} range \citep[e.g.,][]{Hermes2021}. The resulting light curve has an effective central wavelength of $2520$\,\AA{}, and contains 28 datapoints over $\approx 8$ continuous hours. 

\subsection{Spectroscopy}\label{sec:data.spec}

\cite{Shen2018} obtained a low-resolution spectrum ($R \approx 300$) of LP\,398-9 from $3400-9000$\,\AA{} with the Nordic Optical Telescope (NOT). They noted absorption lines indicating the presence of carbon, oxygen, magnesium, and calcium in the stellar photosphere, and measured a radial velocity $20 \pm 60$\,\kms{}. 

We obtained a mid-resolution spectrum of LP\,398-9 on 2020~November~14 using the Dual-Imaging Spectrograph (DIS) on the 3.5-m telescope at the Apache Point Observatory (PI: Chandra). We employed the B1200/R1200 gratings and a 1.5 arcsec slit width, providing a spectral resolution (Gaussian $\sigma$) of 2\,\AA{} on the blue end and 0.5\,\AA{} on the red end, as estimated with sky emission lines. We obtained four 10-minute exposures bracketed by arc lamp exposures to ensure a reliable wavelength calibration. Our data covered a wavelength range of 4000--7000\,\AA{}, with a dichroic gap between 5200--6000\,\AA{}. We performed a standard data reduction---bias correction, flat-fielding, aperture extraction, and wavelength calibration---using the pipeline tools in IRAF \citep{Tody1986}. 

We compared the \ion{O}{1} absorption line at $\lambda 7002.23$\,\AA{} with its theoretical rest wavelength to derive a radial velocity $80 \pm 10$ \kms{}, which is consistent with the earlier measurement made by \cite{Shen2018}. We also used the $\lambda 6300$\,\AA{} \ion{O}{1} sky emission line to verify that our absolute wavelength calibration is accurate to within 5 \kms{}. We therefore do not detect radial velocity variations that would indicate the presence of a close binary companion. As noted by \cite{Shen2018}, the low radial velocity is somewhat unexpected considering LP\,398-9's hypervelocity space motion, but could be a statistical fluctuation due to small numbers and selection effects.

\section{Analysis}\label{sec:analysis}

In this section we analyze our spectro-photometric observations of LP\,398-9.  We confirm the hypervelocity nature of LP\,398-9 with the latest data from Gaia ($\S$\ref{sec:analysis.hypervel}). We model the spectral energy distribution of LP\,398-9 and its excess infrared emission to derive the system's parameters ($\S$\ref{sec:analysis.sed}). We present multi-band light curves that exhibit a 15.4\,hr period, which we interpret as surface rotation ($\S$\ref{sec:analysis.rot}). Finally, we present follow-up spectroscopy that reveals the presence of circumstellar carbon ($\S$\ref{sec:analysis.spec}).

\subsection{Hypervelocity}\label{sec:analysis.hypervel}

We confirm the hypervelocity status of LP\,398-9 using newly released astrometric data from Gaia EDR3 \citep{GaiaCollaboration2021, Lindegren2021, Lindegren2021c}. LP\,398-9 is among the highest tangential velocity stars in the \textit{Gaia} dataset. Furthermore, it is the sole 15-sigma outlier in the tangential velocity distribution of stars within seven degrees of the SN remnant G70.0-21. Previously, \cite{Shen2018} used data from Gaia DR2 to find a 99\% credible interval of $v_T = [700,1500]$ \kms{} for the tangential proper motion velocity of LP\,398-9. The EDR3 parallax is nearly twice as precise as DR2, and the proper motions are more precise by a factor of four. We use the affine-invariant Markov Chain Monte Carlo (MCMC) sampler \texttt{emcee} \citep{Foreman-Mackey2013,Foreman-Mackey2019} to sample the new astrometric data from Gaia EDR3 and consequently derive an updated posterior distribution for $v_T$. We include all covariances and perform a zero-point correction to the parallax following \cite{Lindegren2021c}. We employ the exponentially decreasing space density prior described in \cite{Bailer-Jones2015}, although its effect is likely minimal since the EDR3 parallax is quite precise ($\varpi / \sigma_{\varpi} \sim 18$). We find a median tangential velocity $v_T = 1010 \pm 60$ \kms{}, with a 99\% credible interval of $v_T = [890, 1150]$ \kms{}. The implied total Galactocentric velocity (including our radial velocity measurement from $\S$\ref{sec:data.spec}) is $v_{\rm Gal} = 1130 \pm 60$ \kms{}. This secures LP\,398-9's status as a hypervelocity star, despite its low apparent radial velocity. The broad conclusions from \cite{Shen2018} about LP\,398-9's Galactic trajectory are unchanged.

\subsection{SED Model and IR Excess}\label{sec:analysis.sed}

To investigate the infrared excess of LP\,398-9, we first fit a stellar model to the observed photometry at optical wavelengths, and then compare the predicted stellar flux in the infrared to the observed IR photometry. We fit the SED of LP\,398-9 in the SDSS \textit{ugriz} passbands using a bespoke grid of model spectra computed with the atmosphere code described in \citet{Blouin2018,Blouin2018a}. Since the non-detection of hydrogen lines in the spectrum of LP\,398-9 rules out the presence of atmospheric hydrogen, we assume a helium-dominated atmosphere. The SED can also be altered by the presence of metals due to line blanketing and changes in the atmospheric opacity. We therefore use the NOT spectrum of LP\,398-9 observed by \cite{Shen2018} to set the metallicity to $\text{[Ca/He] = -11.2}$, with other abundances scaled to solar proportions. We adopt a surface gravity $\log{g} = 5.5$ when computing the model spectra, and verify that neither the photometric nor spectroscopic fits are very sensitive to the assumed metallicity and surface gravity.

We assume flat priors on the stellar parameters: $2500 \leq T_{\text{eff}}/K \leq 10000 $ and $0.05 \leq R / R_{\odot} \leq 1$. We adopt a Gaussian prior on the parallax $\varpi/\text{mas}$ with a mean and standard deviation defined by the \textit{Gaia} EDR3 astrometric measurement $\pi = 1.19 \pm 0.06$ mas. To prevent any single band with underestimated uncertainties from dominating the fit, we implement a floor uncertainty of 0.03 magnitudes in all bands \citep[e.g.,][]{Bergeron2019a}. We define a log-likelihood (the $\chi^2$ statistic multiplied by $-0.5$) to compare the model fluxes to the observed fluxes and perform a preliminary fitting step to estimate the atmospheric parameters of LP\,398-9. We maximize the photometric likelihood over the stellar parameters ($ T_{\text{eff}}$ and $R$). Next, using the preliminary stellar parameters as initial values, we sample the posterior distributions of $\teff$, $R$, and $\varpi$ using \texttt{emcee}. The stellar parameters are summarized in Table \ref{tab:params}, along with uncertainties computed by taking the standard deviation of the respective MCMC samples. We adopt the posterior sample with the highest log-likelihood as our best-fitting stellar parameters for LP\,398-9: $T_{\text{eff}} = 7500 \pm 100$ K, $R = 0.20 \pm 0.01\ R_\odot$.

We repeated our analysis assuming oxygen-dominated and carbon-dominated atmospheres. The inferred $T_{\text{eff}}$ and $R$ are similar for the helium and oxygen atmospheres. A carbon-dominated atmosphere is ruled out, since it would require the presence of strong molecular C$_2$ absorption lines that are absent in our spectroscopic observations (discussed further in $\S$\ref{sec:analysis.spec}). The mass of LP\,398-9 is quite uncertain, since the stellar structure is unknown and the star is in a temporarily inflated state. However, given the D$^6$ origin of the system, the mass could plausibly lie in the 0.2--0.8\,$M_\odot$ range \citep{Shen2015,Shen2018}, with the runaway velocity suggesting a mass on the lower end \citep{Bauer2021b}. Assuming a uniform mass prior in this range, our fitted photometric radius implies $\log{g} \approx 5.5 \pm 0.2$. This is consistent with the \ion{Ca}{2} H and K absorption lines on the archival NOT spectrum. 

\begin{deluxetable}{lr}\label{tab:params}
\tablewidth{\columnwidth}
\tablecaption{Parameters of LP\,398-9}
\tablehead{
\colhead{Parameter} & \colhead{Value}}
\startdata
\sidehead{Gaia EDR3}
Source ID & 1798008584396457088 \\
RA (degrees) & 324.61250 \\
Dec. (degrees) & 25.37374 \\
G (mag) & 16.97 \\
BP$-$RP (mag) & 0.41 \\
$\varpi$ (mas) & 1.19 $\pm$ 0.06 \\
Distance (pc) & 840 $\pm$ 40 \\ 
$\mu_{\text{RA}}$ (mas/yr) & 98.28 $\pm$ 0.07 \\
$\mu_{\text{DEC}}$ (mas/yr) & 240.18 $\pm$ 0.06 \\
$v_T$ (\kms{}) & 1010 $\pm$ 60 \\
\tableline
\sidehead{Stellar Model}
$T_{\rm{eff}}$ (K) & 7500 $\pm$ 100 \\
Radius ($R_\odot$) & 0.20 $\pm$ 0.01 \\
Mass ($M_\odot$) & $\rm{U}\,[0.2, 0.8]$ (assumed) \\
$\log{\left[g, \rm{cm}\, \rm{s}^{-2}\right]}$ & 5.5 $\pm$ 0.2 \\
$v_{R}$ (\kms{}) & 80 $\pm$ 10 \\
\tableline
\sidehead{Blackbody Infrared Model}
$T_{\text{bb}}$ (K) & 670 $\pm$ 50\\
$R_{\text{bb}}$ ($R_\odot$) & 5.4 $\pm$ 0.9\\
$M_{\text{dust}}$ ($g$) & $\sim 10^{20}$ \\
\tableline
\sidehead{Flat Disk Infrared Model ($i = 0^\circ$)}
$T_{\text{in}}$ (K) & 1300 $\pm$ 200\\
$T_{\text{out}}$ (K) & $\approx 300$ \\
\enddata
\end{deluxetable}

To quantify and model the infrared excess around LP\,398-9 we use the AllWISE catalog magnitudes in the \textit{W1} and \textit{W2} channels. We convert them to fluxes, and subtract the expected flux of LP\,398-9 at these wavelengths using the stellar model described above. To this excess flux, we fit two models: a blackbody, and a flat disk model. The blackbody is parameterized by the temperature $T_{\text{bb}}$ and a normalization factor that depends on the solid angle subtended on the sky. We use the \textit{Gaia} EDR3 parallax of LP\,398-9 to fix its distance, fitting for the remaining normalization radius $R_{\text{bb}}$ in addition to $T_{\text{bb}}$. 

For the disk model, we use the geometrically thin, optically thick debris disk model of \cite{Jura2003}. With a known distance, this model is further constrained by the stellar temperature $\teff$ and radius $R$, with the free parameters being the two temperatures corresponding to the inner edge ($T_{\text{in}}$) and outer edge ($T_{\text{out}}$) of the disk. The inclination of the disk is difficult to constrain with the available observations, although it is probably low (more face-on) given the brightness of the IR excess compared to the star. Additionally, the temperature of the outer edge of the disk $T_{\text{out}}$ is poorly constrained because observations at longer wavelengths are not available. 

\begin{figure}
    \centering
    \includegraphics[width=\columnwidth]{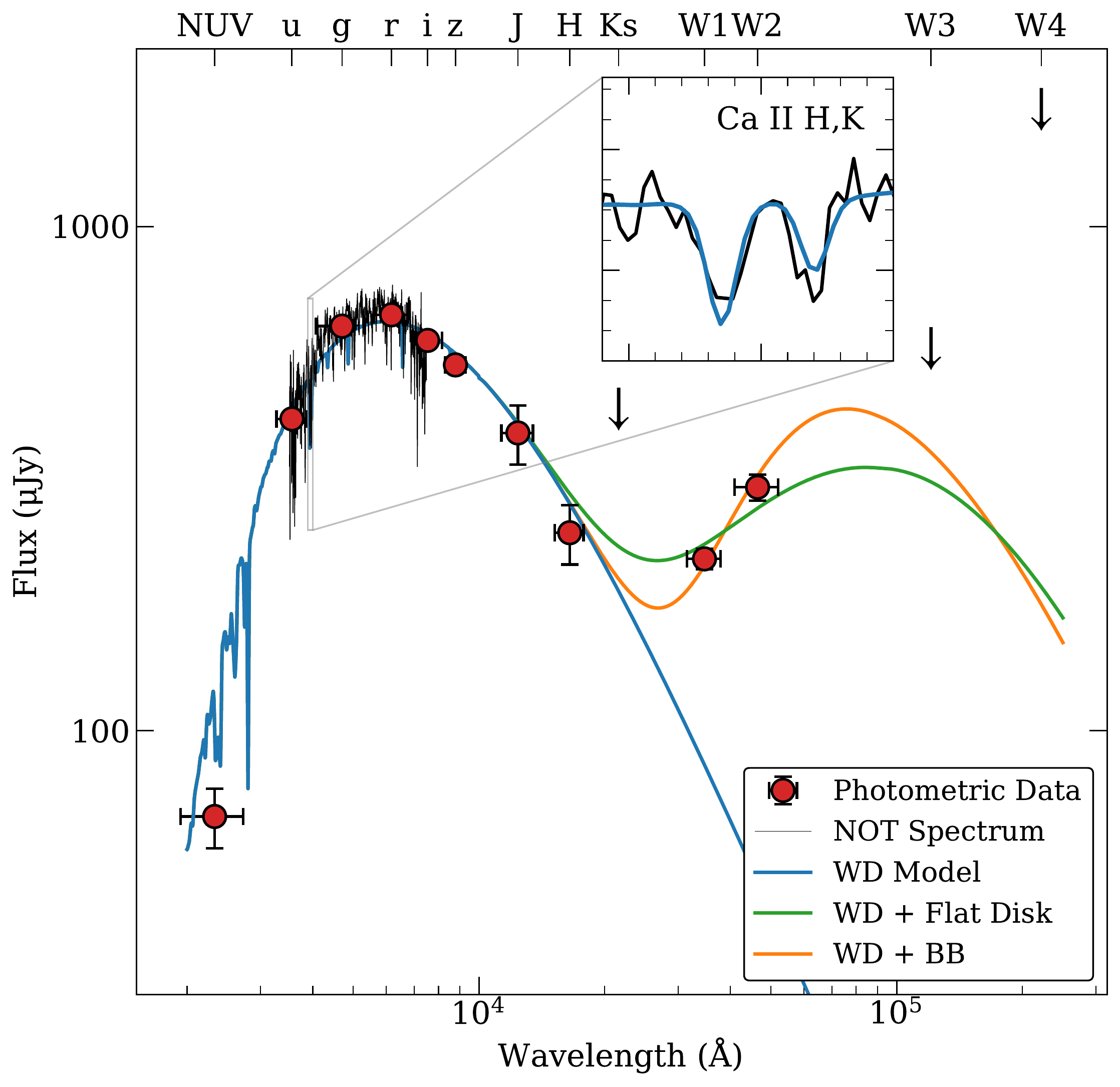}
    \caption{Spectral energy distribution of LP\,398-9 with our best-fitting model for the WD and IR excess. The observed photometric bands are indicated on the top axis. The inset shows the 3890--4000\,\AA{} region of the archival NOT spectrum, with the best-fitting stellar model spectrum overlaid. Downward arrows indicate upper limits. Uncertainties are 1-sigma after applying our 0.03 mag floor, and in some cases are smaller than the marker size.}
    \label{fig:sed_model}
\end{figure}

We fit the respective models using the nonlinear minimization algorithms in \texttt{scipy.optimize} \citep{Virtanen2020}. We repeat the fit on $10^4$ Monte Carlo replicates of the data (adding the relevant Gaussian errors to the observed fluxes) to estimate the uncertainties on the parameters. The pure stellar, stellar + blackbody, and stellar + flat disk models are illustrated with the observed SED in Figure \ref{fig:sed_model}, and the best-fit IR model parameters are summarized in Table \ref{tab:params}. 

The blackbody temperature can be interpreted as the average temperature of an inflated, optically thin dust shell \citep{Xu2018}. Given the low temperature $T_{\text{bb}} \approx 670$\,K and large normalization radius $R_{\text{bb}} \approx 5.4\, R_\odot$, we can rule out stellar or sub-stellar companions. A brown dwarf with $T_{\rm{eff}} \sim 700$\,K would have a radius $\sim 0.1\, R_{\odot}$ \citep[e.g.,][]{Sorahana2013}, almost fifty times smaller than the radius we derive from the normalization of the IR excess. Depending on the typical grain sizes, $T_{\text{bb}} \approx 670$\,K corresponds to dust located at orbital radii $\approx$ 10--30\,$R_\odot$, or around 50--150 times the present radius of the star \citep{Xu2018}. 

The flat disk model does not match the observed fluxes well, indicating that a geometrically thin, optically thick disk is insufficient to describe the material around LP\,398-9. A low inclination ($\approx 0^\circ-30^\circ$) provides the best match to the \textit{W1} and \textit{W2} data. The inclination and inner disk temperature are degenerate -- assumed inclinations of $[0^\circ, 30^\circ, 60^\circ]$ result in best-fit $T_{\text{in}} \approx [1300, 1600, 2000]$\,K. Depending on the size and composition of the grains, circumstellar dust sublimates at temperatures between 1200-- 2000\,K, and consequently any material inwards of the radius corresponding to $T_{\text{in}}$ rapidly sublimates away, leaving a gap between the star and the debris disk  \citep{vonHippel2007,Rafikov2012,Xu2018}. Therefore, the inferred inner temperature $T_{\text{in}} \approx 1300$\,K is broadly consistent with dust sublimation temperatures, although the flat disk geometry seems unlikely. Our favored interpretation is a circumstellar dust shell model with a characteristic temperature of $T_{\text{bb}} \approx 670$\,K. 

\subsection{Periodic Photometric Variability}\label{sec:analysis.rot}

We searched the LT photometry of LP\,398-9 for periodic variations using a Lomb-Scargle periodogram \citep{Lomb1976, Scargle1982}. We find a dominant peak at $f = 1.55~\text{day}^{-1}$, corresponding to a period of $\approx 15.4$ hours. We repeat our periodogram analysis on the ZTF photometry. Fitting the entire ZTF baseline from 2018~April~20 to 2021~April~27 yields a period $1\%$ smaller than the LT period. However, this ZTF light curve has significant long-term brightness variations that contaminate the periodogram, requiring pre-whitening to remove trends. If we instead fit only the most recent ZTF campaign from 2020~January~6 to 2021~April~27 without pre-whitening, we derive a period within $0.03\%$ of the LT period. We therefore use only the ZTF data from 2020~January~6 onwards in our subsequent analysis. 

\begin{figure}
    \centering
    \includegraphics[width=\columnwidth]{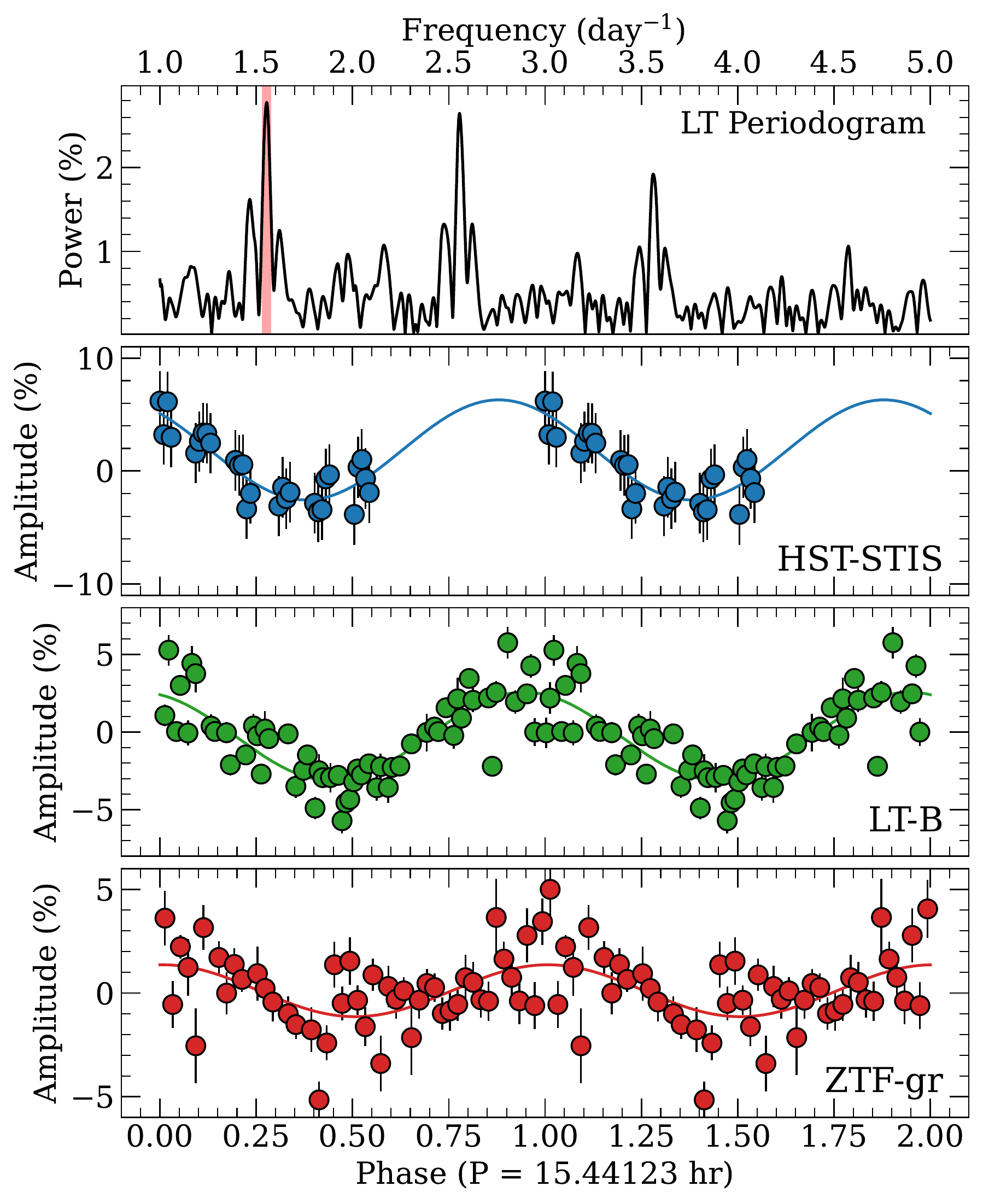}
    \caption{Lomb-Scargle periodogram of the LT light curve of LP\,398-9, along with the phase-folded HST-STIS, LT-$B$, and ZTF-$gr$ light curves. The STIS datapoints are unbinned and consecutive from a single epoch, while the LT and ZTF datapoints are binned in phase. We overlay the sinusoidal models used to estimate the variability amplitude of each light curve. }
    \label{fig:pp_lc}
\end{figure}

We adopt the peak of the LT periodogram as our assumed period. To estimate the uncertainty of this measurement, we computed 2500 Monte Carlo replicates of the LT light curve by adding Gaussian uncertainties, and computed half the difference between the 84th and 16th quantiles of the resulting period distribution. We derive a best-fit period $P = 15.441 \pm 0.016$\,hr. We phase-fold the STIS, LT, and ZTF light curves and display them in Figure \ref{fig:pp_lc}. We fit a sinusoidal model to each phase-folded (un-binned) light curve to estimate the variability amplitudes. We find amplitudes of $4.4 \pm 0.6\,\%$ for STIS, $2.8 \pm 0.2\,\%$ for LT-$B$, and $1.3 \pm 0.3\,\%$ for ZTF-$gr$. Figure \ref{fig:pp_lc} shows a slight color-dependent phase difference between the three datasets. However, simply changing the assumed period by as little as 0.3\% (to $P =  15.40317$, for example) can erase this apparent phase difference. Therefore, the most plausible explanation is that this phase difference is non-physical, and is simply an artifact of systematic uncertainties in our photometry and period measurement. 

\begin{figure}
    \centering
    \includegraphics[width=\columnwidth]{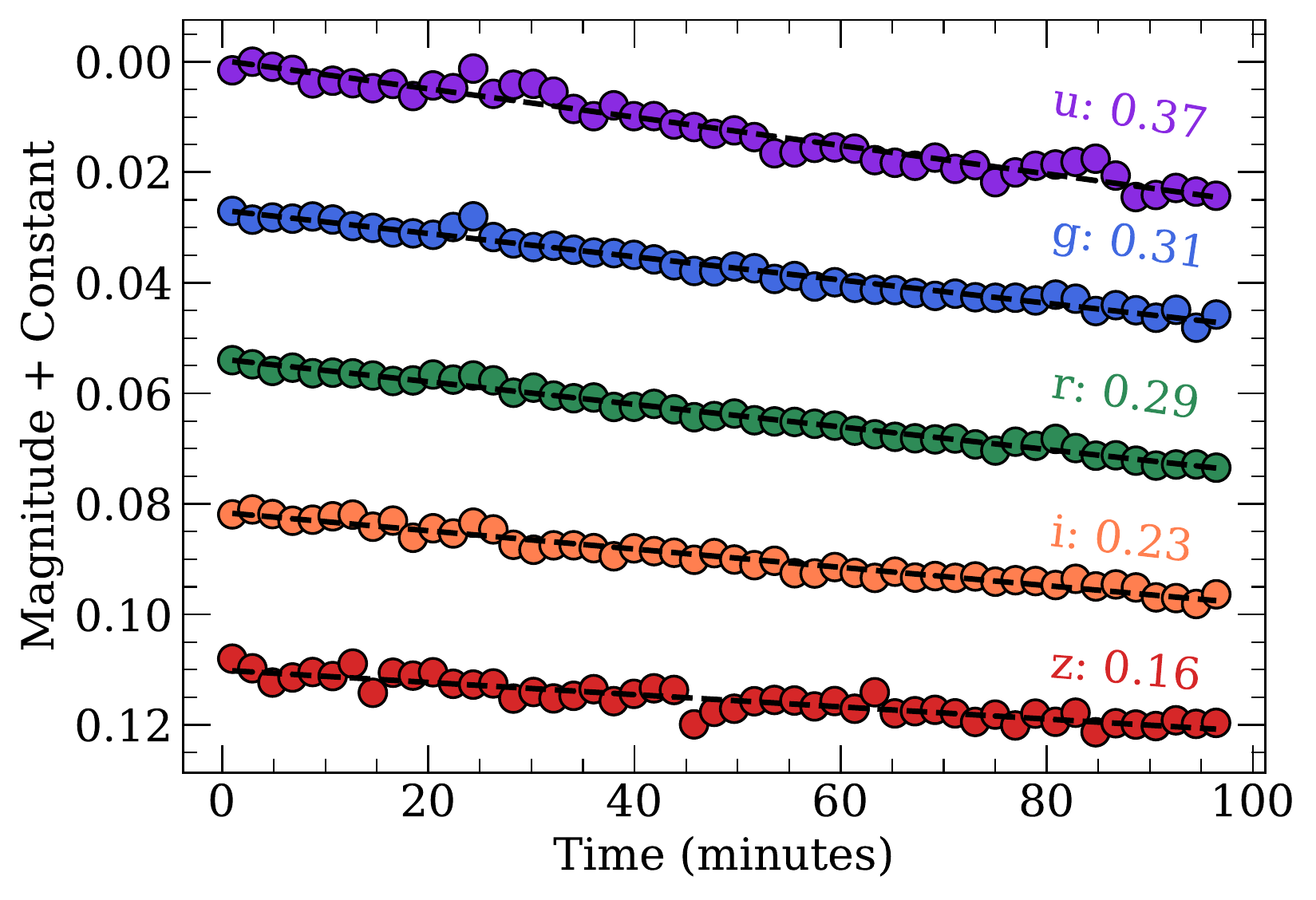}
    \caption{100-minute HiPERCAM light curve of LP\,398-9. The $ugriz$ bands are arranged from top to bottom. We overlay a linear fit to each dataset with a black dashed line. The color-dependent trend is visually discernible, with the bluest $u$ band declining twice as fast as the reddest $z$ band. For each band we show the fitted slope of the flux decline in magnitudes per day.}
    \label{fig:hipercam}
\end{figure}

Our 100-minute HiPERCAM $ugriz$ observations are illustrated in Figure~\ref{fig:hipercam}. The time baseline of this observation covers a fraction of our derived rotational period, and therefore represents a quasi-linear segment of the rotationally modulated flux variation. Both our HiPERCAM-$ugriz$ and subsequent INT-$g$ observations (not shown here) caught LP\,398-9 in the dimming phase. We quantify the color dependence of the flux decline by fitting a simple linear regression to the light curves in each band using the trimmed least squares approach described in \cite{Cappellari2013a}. The rate of decline (slope) is strongly wavelength-dependent, with the bluest photometry declining twice as quickly as the reddest. The $g$-band slope is identical within uncertainties between the HiPERCAM and INT data. The trend of the HiPERCAM color-dependence (blue declining faster than red) is consistent with the trend in the variability amplitudes of the STIS, LT-$B$, and ZTF-$gr$ light curves. 

The color-dependence of the photometric variability could have several plausible explanations. A convective star spot with a temperature lower than the surrounding photosphere could produce such a signal, although it is challenging to constrain the temperature and size of the spot \citep[e.g.,][]{Pelisoli2021a}. Alternatively, if a portion of the stellar surface is polluted with metals, there could be an opacity or line blanketing contrast as the star rotates. Additionally, light filtering through the circumstellar dust could be reddened by extinction. 

\begin{figure}
    \centering
    \includegraphics[width=\columnwidth]{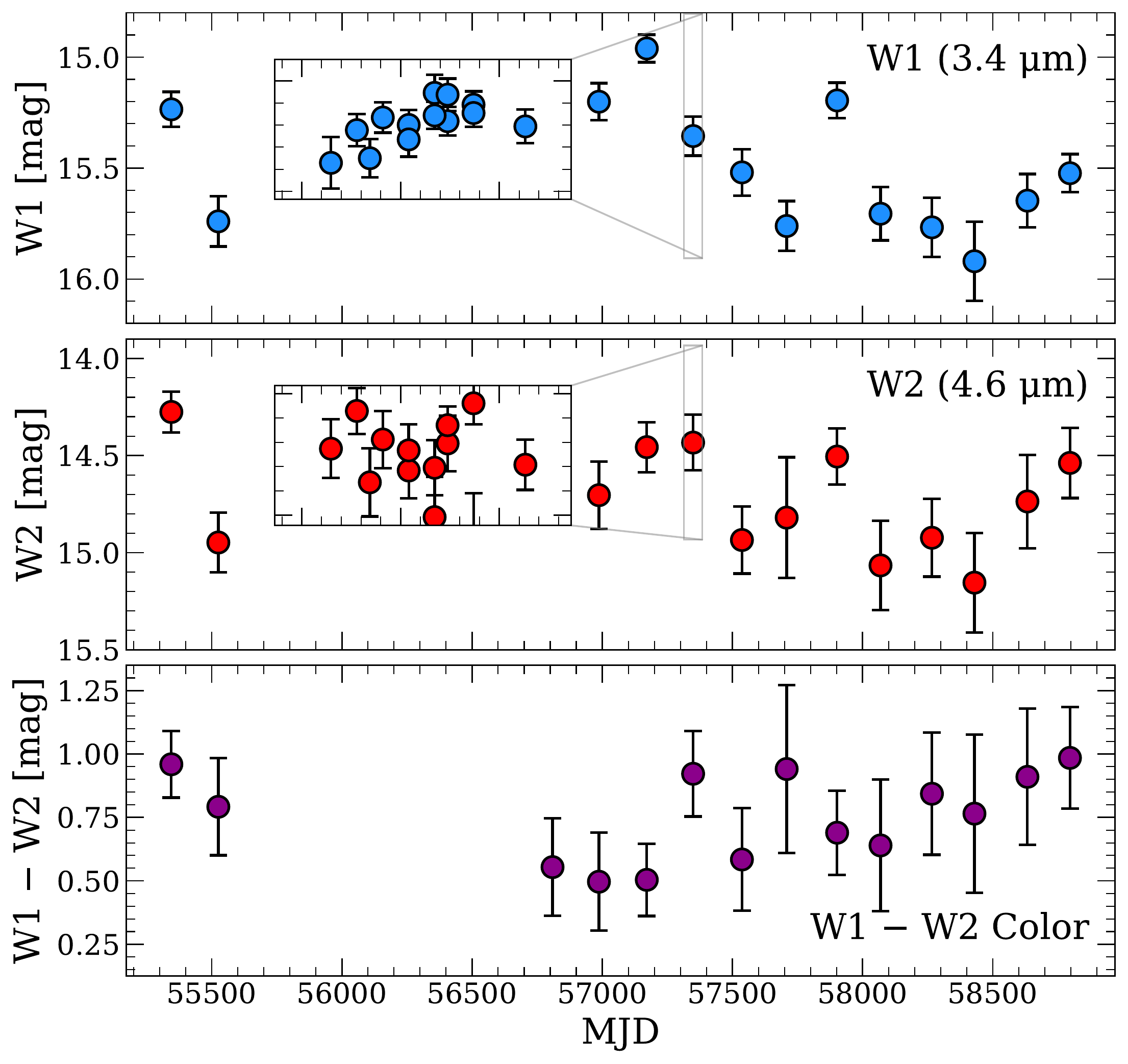}
    \caption{Long-term light curves from the duration of the \textit{WISE} mission in the \textit{W1} and \textit{W2} bands, along with their difference. We average the individual measurements from each observing run to produce these datapoints. The insets show individual measurements from a single \textit{WISE} observing run. The insets vertically span 1 magnitude, and horizontally span 2 days.}
    \label{fig:wise_lc}
\end{figure}

As described in $\S$\ref{sec:data.phot}, we compiled a \textit{WISE} light curve of LP\,398-9 spanning 9 years, grouped into 14 approximately day-long observing runs spaced 180 days apart. Phase-folding the \textit{WISE} light curve to the inferred stellar rotation period of 15.4\,hr does not produce any coherent signal. This is expected, since the IR flux is dominated by emission from the surrounding circumstellar dust, which would greatly dilute any photometric variability from the star itself. However, the \textit{WISE} data of LP\,398-9 is overall more variable than the photometric uncertainties would indicate. We average the measurements in each observing run and illustrate the long-term \textit{W1} and \textit{W2} fluxes along with the \textit{W1}$-$\textit{W2} color in Figure \ref{fig:wise_lc}. There is significant variation in both the \textit{W1} and \textit{W2} fluxes. The \textit{W1}$-$\textit{W2} color marginally varies as well, potentially indicating a varying surface area or temperature of the circumstellar material. Recent evidence has shown that many WDs with debris disks are IR-variable \citep{Swan2019a,Swan2020}. Those systems are thought to exhibit stochastic variation over long timescales due to collisional dust production and depletion \citep[e.g.,][]{Kenyon2017,Kenyon2017b,Swan2021}. 

\subsection{Spectroscopic Evidence of Circumstellar Carbon}\label{sec:analysis.spec}

\begin{figure}
    \centering
    \includegraphics[width=\columnwidth]{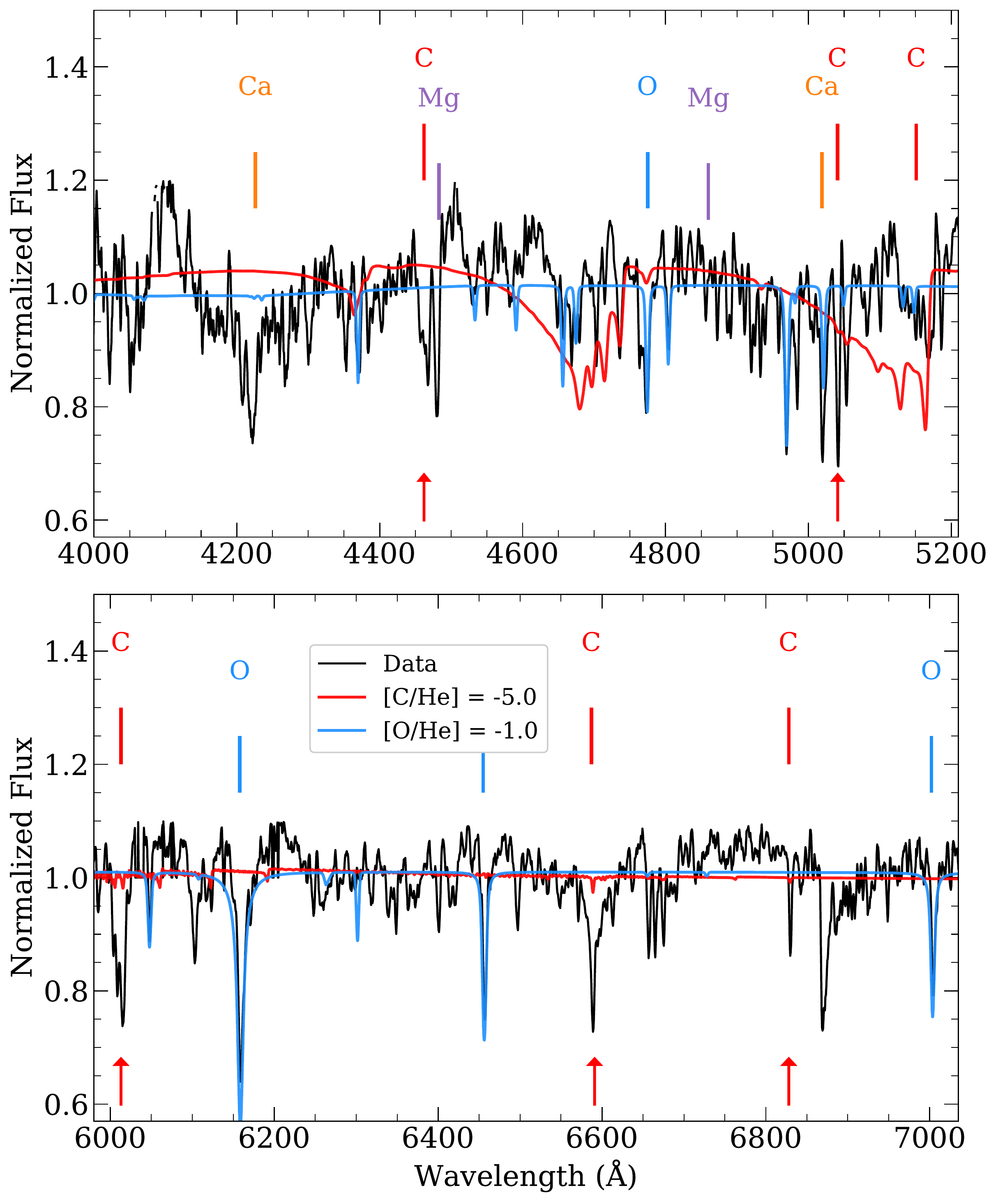}
    \caption{Mid-resolution spectrum of LP\,398-9 from the 3.5-m telescope at the Apache Point Observatory, split into two wavelength regions. The spectrum is smoothed with a $3$\,\AA\ boxcar. We indicate identified absorption lines of carbon, oxygen, magnesium, and calcium. We overlay in red a theoretical spectrum with the same stellar parameters as LP\,398-9 and an atmospheric carbon abundance of [C/He] = -5.0, and independently in blue an oxygen abundance of [O/He] = -1.0. Red arrows indicate carbon absorption lines that cannot be explained by the stellar model, suggesting that they are formed in circumstellar material instead.}
    \label{fig:apo}
\end{figure}

In Figure \ref{fig:apo} we illustrate the APO spectrum of LP\,398-9, with several identified absorption lines of carbon, oxygen, magnesium, and calcium highlighted. We characterize the spectrum using theoretical models described in \cite{Blouin2018,Blouin2018a}. We adopt the convention that [X/He] denotes the logarithm of the number density ratio between species X and He. We fix the stellar parameters to the values computed from the SED fit in $\S$\ref{sec:analysis.sed} and compute single-abundance spectra over a range of C and O to investigate the atmospheric composition We overlay in red a model spectrum with $\text{[C/He]} = -5.0$, and in blue a model with $\text{[O/He]} = -1.0$. The short-wavelength end of the spectrum demonstrates that this carbon abundance is already pushing the higher plausible limit, due to the development of strong C$_2$ bands that are absent from our observed spectrum \citep{Swan1857,Johnson1927}. Yet, several narrow carbon lines remain unexplained by the stellar models (red arrows in Figure \ref{fig:apo}). 

The most plausible explanation for this non-photospheric absorption is the presence of circumstellar carbon gas around LP\,398-9. Compared to the stellar photosphere, the circumstellar region has much lower pressures, and consequently contains less molecular C$_2$. This means more atomic C can be present to produce narrow absorption lines without creating large molecular bands on the spectrum. The two most prominent atomic carbon features are \ion{C}{1} lines at 6587.6 \AA\ and 6828.2 \AA\ respectively. Based on their equivalent widths --- 1.9 \AA\ and 0.6 \AA\ respectively --- we can infer a column density of $\sim 10^{14}\ \text{cm}^{-2}$ in the optically thin limit \citep{Johansson1966,Luo1989,Hibbert1993,Draine2011,NIST_ASD}. For comparison, this estimate is between one and two orders of magnitude lower than neutral carbon column densities around typical mass-losing asymptotic giant branch stars \citep[e.g.,][]{Keene1993,Hasegawa2003}. If the gas is co-spatial with the dust and occupies a region of order $R_{\text{bb}}$ from $\S$\ref{sec:analysis.sed}, then the implied gas mass would be similar to the inferred dust mass. Our present data do not constrain the spatial extent or geometric configuration of the gas and dust.  We cannot rule out that a larger reservoir of dust surrounds LP\,398-9, with only the material close to the star being heated and revealed by our observations of the IR excess and \ion{C}{1} absorption.

Although we only report the detection of circumstellar carbon with our current data, future spectroscopy at higher resolution and higher signal-to-noise could reveal absorption from other species. We emphasize that the carbon and oxygen abundances stated in Figure~\ref{fig:apo} are qualitative approximations used to investigate the presence of circumstellar absorption. A more sophisticated stellar model is required to separate the photospheric and circumstellar contributions to the spectrum, and enable a detailed abundance analysis of both components.
 
\section{Results}\label{sec:results}
 
\subsection{Photometric Period}

LP\,398-9 exhibits sinusoidal photometric variation with a period $P \approx 15.4$\,hr. A stellar companion is quite unlikely, given the non-detection of radial velocity variations $\gtrsim 10$ \kms{}. The most natural explanation is that this photometric period corresponds to the rotational period of the star \citep[e.g.,][]{Brinkworth2005,Hermes2021}. There are several possible mechanisms that can produce rotationally modulated variability. WDs with exceptionally strong magnetic fields can have continuum opacity variations on their surfaces, producing photometric variations at the rotational period \citep[e.g.,][]{Ferrario1997,Tremblay2015a}. However, the optical spectrum of LP 398-9 does not indicate a strong magnetic field, given the absence of visible Zeeman splitting. An alternative possibility is the formation of weakly magnetized starspots in the convective envelope of the star \citep[e.g.,][]{Brinkworth2004,Gansicke2020a}. These spots would be cooler than the surrounding regions, producing chromatic variations in the observed flux as they rotated in and out of view \citep[e.g.,][]{Kilic2015a,Reding2020}. 

Another interesting possibility is that the surface inhomogeneity of LP\,398-9 is related to its origin as a D$^6$ SN\,Ia donor. When its companion exploded $\sim 10^5$ years ago, LP\,398-9 would have been blasted by a wave of ejecta. These radioactive ejecta would have been asymmetrically deposited on the surface of LP\,398-9 \citep[e.g.,][]{Bauer2019}. In regular WDs, horizontal mixing can homogenize the surface distribution of pollutants on timescales of $10^1-10^6$~years depending on the temperature and atmospheric composition \citep{Cunningham2021}. Since the stellar structure of LP 398-9 is quite uncertain, it is difficult to estimate the timescale with which diffusion is expected to homogenize the stellar surface. The vertical diffusion timescale is certainly long compared to regular WDs, since LP\,398-9 is inflated to $\approx 10$ times the typical WD radius. If the deposited surface asymmetry does survive to the present day, then it could produce variable line blanketing and continuum opacity, possibly producing the observed photometric signal \citep[e.g.,][]{Brinkworth2004,Kilic2015a,Maoz2015}. If the surface abundances vary along the line of sight, this may be detectable with future time-resolved optical spectra. 

Apart from the origin of the surface inhomogeneity, the rotational period of LP\,398-9 can also be linked to the D$^6$ mechanism. Immediately prior to the SN explosion of the massive companion, the binary system would have been in an extremely compact, mass-transferring stage. The orbital velocity $v_{\rm orb}$ of the donor (mass $M_2$) around the accreting primary (mass $M_1$) is given by
\begin{equation}
    v_{\rm orb}^2 = \frac{G M_1}{a(1+M_2/M_1)}~,
\end{equation}%
where $a$ is the orbital separation. LP\,398-9 would have been ejected from the binary at $v_{\text{orb}}$ when its companion exploded. The amount of low-velocity material left behind by the exploded companion is negligible, so LP\,398-9 has probably not slowed down much since the supernova. This allows us to express the orbital period in terms of the measured space velocity as
\begin{equation}
    P_{\text{orb}} = \frac{2 \pi a}{\left( 1 + M_2/M_1 \right) v_{\text{orb}}}
    = \frac{2 \pi G M_1}{(1+M_2/M_1)^2v_{\rm orb}^3}~.
\end{equation}
Assuming that the primary star that exploded as an SN is $\approx 1.0\,M_\odot$, and that the donor mass is somewhere in the range $M_2 = 0.2$--$0.8\,M_\odot$, this gives orbital periods at the time of SN detonation around $P_{\rm orb}=3$--$7\,\rm min$, based on the $\approx 1100$~\kms{} space velocity of LP\,398-9.

Although the tidal quality factor of WDs is uncertain, it is reasonable to expect a degree of tidal synchronization between the spin and orbit of the WDs \citep[e.g.,][]{Iben1998, Fuller2011, Fuller2012, Fuller2012a, Fuller2014, Yu2020tides, Yu2021tides}. There are a handful of known tidally-distorted WDs in compact binaries, and the non-detection of significant tidal heating could imply that synchronization is near-perfect \citep{Piro2011,Benacquista2011}. Under the assumption that tidal locking keeps $P_{\text{rot}} \approx P_{\text{orb}}$, our reasoning implies a rotation period $\approx 3$--7\,minutes when the companion exploded. If LP\,398-9 subsequently conserved its angular momentum, then the initial post-SN rotational period $P_i$ and radius $R_i$ and present-day $P_f$, $R_f$ can be related with %
\begin{equation}\label{angular_p}
    \frac{P_f}{P_i} = \left( \frac{R_f}{R_i} \right)^2
\end{equation}%
We can set $R_i \approx 0.01$--0.02\,$R_\odot$ using the mass-radius relation for $0.2$--$0.8\,M_\odot$ WDs, and set $R_f = 0.20\,R_\odot$ from our SED fit in $\S$\ref{sec:analysis.sed}. Applying Equation \ref{angular_p}, post-SN rotational periods $P_{i} \approx 3$--7\,minutes correspond to present-day rotational periods $P_f \approx 5$--50\,hours, which brackets our observed photometric period of 15.4\,hr for LP\,398-9 ($\S$\ref{sec:data.lc}). A larger initial donor radius (e.g., due to heating during mass transfer) corresponds to a shorter present-day rotational period. Conversely, angular momentum was probably not entirely conserved. Additional effects like magnetic braking and mass loss could have further slowed the rotational rate of the system over time. These rough estimates demonstrate that the observed rotational period can be plausibly linked to the D$^6$ scenario, assuming LP\,398-9 mostly conserved its angular momentum after the SN\,Ia explosion. 

We emphasize here the differences and similarities between LP\,398-9 and GD\,492, another runaway star that was recently found to rotate with an 8.9\,hr period \citep{Hermes2021}. GD\,492 has a peculiar composition that suggests it is the partially burnt accretor left over from a Type Iax supernova \citep{Vennes2017,Raddi2018,Raddi2018a,Raddi2019}. GD\,492 also has a lower space velocity $\simeq 850$\,\kms{}, implying that its donor companion was an He-burning subdwarf rather than another white dwarf \citep{Bauer2019}. In this scenario, GD\,492 is rotating too slowly to be the runaway subdwarf donor, and is most likely the burnt remnant of the accreting primary that underwent a failed supernova. Conversely, the higher space velocity of LP\,398-9 --- and its remarkable spectroscopic similarity to two other D$^6$ systems with space velocities $\gtrsim 1500$\,\kms{} --- supports a D$^6$ origin for this system. In this context, our detected rotation period could point to LP\,398-9 being a D$^6$ donor. 

\subsection{Circumstellar Material}

LP\,398-9 has a strong IR excess that indicates the presence of significant quantities of circumstellar dust. Our optical spectrum shows narrow atomic carbon lines unexplained by the photospheric model, pointing to a circumstellar source of carbon. Our observations suggest a carbon-rich shell of circumstellar material inflated to more than an order of magnitude larger than present-day radius of the star.

The presence of carbon-rich circumstellar material can be plausibly explained by the D$^6$ origin of LP\,398-9. The SN\,Ia explosion of its stellar companion would have significantly polluted the atmosphere of LP\,398-9, depositing thermal and radioactive energy and causing the atmosphere (dominated by carbon and oxygen) to puff up. The radius of LP\,398-9 inferred from its luminosity and parallax is an order of magnitude larger than that of a typical WD, suggesting that it remains in a somewhat puffed phase like the other two runaways found in \cite{Shen2018}. Mechanical energy from the companion's SN\,Ia explosion could also have pushed the circumstellar material outwards to larger orbital radii. If some fraction of the inflated layers had detached from the star, it would appear today as an extended carbon-rich circumstellar shell. A low-velocity tail of the SNIa ejecta could also have been retained by LP\,398-9 after the explosion. A promising avenue to resolve this question is performing 3-D simulations of the D$^6$ scenario, for example to ascertain the amount and composition of SN ejecta that remain bound to the donor \citep[e.g.,][]{Tanikawa2018}.

Under the assumption of an optically thin dust shell surrounding LP\,398-9, we can use our observations to approximate the total dust mass. We adopt the fitted black-body dust temperature and Gaia-inferred distance from Table \ref{tab:params}. We assume a dust opacity $\kappa(3.4\,\micron) = 500$ cm$^2$\,g$^{-1}$ \citep{Draine1984, Draine2003, Woitke2016}, consistent with amorphous carbon grains. We estimate the dust mass using the relation from \citet{Hildebrand1983}: %
\begin{equation}
     M_{\rm{dust}} = \frac{F(\lambda) D^2}{\kappa(\lambda) B(\lambda, T_d)}
\end{equation} %
Here $B(\lambda, T_d)$ is the Planck function, and $D$ is the parallax-inferred distance. Substituting the \textit{WISE} \textit{W1} flux as $F(\lambda)$ provides an order-of-magnitude total dust mass $M_{\rm dust} \approx 10^{20}\, g = 10^{-13}\, M_\odot$. This dust mass is comparable to that of known dust disks around WDs \citep{Jura2003,Reach2005}, although we stress that LP\,398-9's origin and composition is quite different from those systems. 

\section{Discussion}\label{sec:discussion}

We have presented two new lines of evidence tying LP\,398-9 to the D$^6$ SNIa progenitor scenario: circumstellar material and surface rotation. We interpret the 15.4\,hr photometric signal as a signature of rotationally modulated brightness variations, potentially stemming from surface inhomogeneities left over from the SN explosion $\sim 10^5$~years ago. The rotational period itself can be explained by angular momentum conservation of LP\,398-9 after its companion exploded, assuming the binary was tidally locked at the point of detonation. 

Several effects could have altered the rotational rate in the time since the supernova. Since tidal dissipation primarily occurs near the surface of the WD, the surface layers could preferentially synchronize without fully spinning up the core, invalidating our assumption of a rigid-body rotation \citep[e.g.,][]{Goldreich1989,Fuller2012,Fuller2012a,Fuller2014}. Additionally, LP\,398-9 could have lost angular momentum in the time since the SN explosion. An obvious culprit could be the inflated stellar layers that today present as circumstellar dust tens of stellar radii away. The present-day total dust mass is a minuscule fraction of the stellar mass, making it an unlikely route for angular momentum to leave the system. However, there could have been episodes of further mass loss in the time since the supernova. 

Although it is uncertain precisely what processes cause the D$^6$ donor WDs to inflate to their presently observed radii, this inflated phase is probably only temporary \citep{Shen2018,Bauer2019}. Due to selection effects, we are most likely to find D$^6$ donors in their inflated state, since these are detectable out to a larger search volume. Our fitted stellar parameters imply a Kelvin-Helmholtz timescale $\tau_{\text{KH}} \gtrsim 20$\,Myr for LP\,398-9, much longer than the flight time $\approx 0.1$\,Myr from the SN remnant G70.0-21.5. As LP\,398-9 radiates away energy deposited by tidal heating and the SN itself, it will contract in size and gradually return to the WD cooling track. In the absence of any mechanism to remove angular momentum from the system, its rotation will speed up as it shrinks. Therefore, a plausible long-term outcome of the D$^6$ scenario is a class of rapidly rotating hypervelocity white dwarfs. These might have helium-dominated or carbon-dominated spectral types. However, such hypervelocity systems leave the Galaxy on short timescale $\lesssim 10$~Myr, making it quite unlikely that we could detect them. Due to these selection effects, we are overwhelmingly likely to find D$^6$ donors in their inflated state.

In ordinary WDs with infrared-emitting disks, the lifetime of dust in the region where it produces infrared excess is limited by Poynting-Robertson drag \citep{Rafikov2011}. However, the Poynting-Robertson timescale for LP\,398-9 is much longer than typical WDs, largely due to its inflated state. Another important effect is sub-micron dust removal via radiation pressure, where the sub-micron dust is created via collisions \citep{Chen2020}. The collisional lifetime can be high if the infrared excess is produced by a dynamically cold disk with a small internal velocity dispersion, which is nominally consistent with the narrow observed lines of circumstellar carbon ($\S$\ref{sec:analysis.spec}). However, given the poor spectral resolution of our data, this is not a strong constraint. If the collisional lifetime is indeed short, a continuous dust production mechanism could be required to explain the infrared excess $\sim 10^5$ years after the formation of the system. Ongoing dust production would also account for the IR flux variability we describe in $\S$\ref{sec:data.phot}.

When a normal isolated WD has an observed IR excess, the usual explanation is circumstellar dust from a disrupted planetesimal \citep[e.g.,][]{Debes2002,Jura2003,Farihi2016,Veras2021review}. While the existence of disrupted planetesimals around LP\,398-9 is tantalizing, it is quite unlikely. The D$^6$ origin of LP\,398-9 implies that it was until recently ($\sim 10^5$ years ago) in a close binary system with another WD. During the final stages of binary evolution, the two WDs in the theorized D$^6$ scenario would be orbiting too close for a planet to maintain a stable orbit around LP\,398-9. While it is plausible that the stars possessed a circumbinary planetary system, any surviving planetesimals must have occupied orbits wide enough to avoid engulfment during giant branch evolution, yet close enough to remain bound to LP\,398-9 after the SN\,Ia of its companion. This scenario can be tested with future infrared spectroscopy --- for example with the James Webb Space Telescope \citep{Gardner2006} --- that would reveal the temperature profile, composition, and geometry of the circumstellar material. These observations could also be used to search for thermonuclear ashes deposited by the past supernova.

One question that remains is why LP\,398-9 (`D6-2') is the only D$^6$ candidate with an infrared excess, out of the three candidates found by \cite{Shen2018}. `D6-1' has secure \textit{WISE} data with no detectable IR excess (Figure \ref{fig:cmd}), while `D6-3' does not have secure data in \textit{WISE} due to a crowded field. One possible explanation is that the other candidates are the products of older supernovae, and could have consequently lost their circumstellar shell over time. LP\,398-9 is the only D$^6$ star with an associated SN remnant, suggesting that it was ejected recently enough that the remnant did not dissipate into the interstellar medium. Conversely, all three D$^6$ stars are clustered closely together on the color-magnitude diagram and have similar low-resolution spectra, perhaps indicating that they are in similar stages of their evolution. D6-1 and D6-3 should be monitored to search for rotationally modulated variations. If they are indeed the product of older supernovae that LP\,398-9, then their surface composition may have homogenized enough to make their rotational signal undetectable. Alternatively, if stark differences persist between LP\,398-9 and other D$^6$ donors, then it is important to consider alternative hypotheses for LP\,398-9's origin \citep{Bauer2021b,Neunteufel2021}.

Future observations of both LP\,398-9 and the other D$^6$ candidates could aid in resolving these questions by placing firmer constraints on the time elapsed since the respective SNe\,Ia. Further spectroscopic observations of other D$^6$ candidates could search for carbon absorption indicative of circumstellar dust that is too cool to produce an IR excess in \textit{WISE}. Spectroscopy at higher resolution and signal-to-noise across a wide wavelength range could also search for absorption by other elements in the circumstellar material. Additionally, more theoretical modelling could estimate the timescales over which ejected runaways cool and return to the WD cooling track. Understanding these unique systems will shed light on the mechanism and after-effects of the double-degenerate channel for SNe\,Ia. 

\section*{Data Availability}

The broadband photometry, \textit{WISE} light curve, ZTF light curve, HiPERCAM light curve, and \textit{HST}-STIS data are all available from respective public data archives. The corresponding author will gladly share any other data products and analysis code upon request. 

\acknowledgments

We thank the anonymous referee for detailed comments that improved the paper. VC thanks Turner Woody for helpful conversations. VC acknowledges support from the James Mills Peirce fellowship at Harvard University, and the Institute for Data-Intensive Engineering and Science at Johns Hopkins University. VC and HCH were supported in part by Space\,@\,Hopkins. VC, HCH, and NLZ were supported in part by NASA-ADAP 80NSSC19K0581. NLZ acknowledges support from J. Robert Oppenheimer Visiting Professorship and the Bershadsky Fund at the Institute for Advanced Study. SB acknowledges support from the Laboratory Directed Research and Development program of Los Alamos National Laboratory (20190624PRD2) and from the Banting Postdoctoral Fellowships program, administered by the Government of Canada. AS acknowledges support from STFC grant ST/R000476/1. KJS is supported by NASA through the Astrophysics Theory Program (NNX17AG28G). The design and construction of HiPERCAM was funded by the European Research Council under the European Union’s Seventh Framework Programme (FP/2007-2013) under ERC-2013-ADG Grant Agreement no. 340040 (HiPERCAM). VSD and HiPERCAM operations are supported by STFC grant ST/V000853/1. 

Based in part on observations obtained with the Apache Point Observatory 3.5-m telescope, which is owned and operated by the Astrophysical Research Consortium. Based in part on observations made with the Gran Telescopio Canarias (GTC), installed in the Spanish Observatorio del Roque de los Muchachos of the Instituto de Astrofísica de Canarias, in the island of La Palma. Based in part on observations obtained with the Isaac Newton Telescope, operated on the island of La Palma by the Isaac Newton Group of Telescopes in the Spanish Observatorio del Roque de los Muchachos of the Instituto de Astrofísica de Canarias. The Liverpool Telescope is operated on the island of La Palma by Liverpool John Moores University in the Spanish Observatorio del Roque de los Muchachos of the Instituto de Astrofisica de Canarias with financial support from the UK Science and Technology Facilities Council. This work has made use of observations made with the NASA/ESA {\em Hubble Space Telescope}, obtained from the Data Archive at the Space Telescope Science Institute, which is operated by the Association of Universities for Research in Astronomy, Inc., under NASA contract NAS 5-26555. These observations are associated with program \#15871, and funding in-part provided by programs \#15871 and \#15918.

This work has made use of data from the European Space Agency (ESA) mission {\it Gaia} (\url{https://www.cosmos.esa.int/gaia}), processed by the {\it Gaia} Data Processing and Analysis Consortium (DPAC, \url{https://www.cosmos.esa.int/web/gaia/dpac/consortium}). Funding for the DPAC has been provided by national institutions, in particular the institutions participating in the {\it Gaia} Multilateral Agreement. We gratefully acknowledge NASA’s support for construction, operation, and science analysis for the GALEX mission, developed in cooperation with the Centre National d’Etudes Spatiales of France and the Korean Ministry of Science and Technology. This publication makes use of data products from the Wide-field Infrared Survey Explorer, which is a joint project of the University of California, Los Angeles, and the Jet Propulsion Laboratory/California Institute of Technology, funded by the National Aeronautics and Space Administration. This publication also makes use of data products from NEOWISE, which is a project of the Jet Propulsion Laboratory/California Institute of Technology, funded by the Planetary Science Division of the National Aeronautics and Space Administration. This publication makes use of data products from the Two Micron All Sky Survey, which is a joint project of the University of Massachusetts and the Infrared Processing and Analysis Center/California Institute of Technology, funded by the National Aeronautics and Space Administration and the National Science Foundation. This research has made use of the NASA/IPAC Infrared Science Archive, which is funded by the National Aeronautics and Space Administration and operated by the California Institute of Technology. This research has made use of the VizieR catalogue access tool, CDS, Strasbourg, France. The original description of the VizieR service was published in 2000, A\&AS 143, 23. This research has made extensive use of NASA's Astrophysics Data System Bibliographic Services. Based on observations obtained with the Samuel Oschin 48-inch Telescope at the Palomar Observatory as part of the Zwicky Transient Facility project. ZTF is supported by the National Science Foundation under Grant No. AST-1440341 and a collaboration including Caltech, IPAC, the Weizmann Institute for Science, the Oskar Klein Center at Stockholm University, the University of Maryland, the University of Washington, Deutsches Elektronen-Synchrotron and Humboldt University, Los Alamos National Laboratories, the TANGO Consortium of Taiwan, the University of Wisconsin at Milwaukee, and Lawrence Berkeley National Laboratories. Operations are conducted by COO, IPAC, and UW. Parts of the results in this work make use of the colormaps in the CMasher package. 

\software{\texttt{numpy} \citep{Harris2020}, 
\texttt{scipy} \citep{Virtanen2020}, 
\texttt{emcee} \citep{Foreman-Mackey2013,Foreman-Mackey2019}, 
\texttt{matplotlib} \citep{Hunter2007}, 
\texttt{lmfit} \citep{Newville2014}, 
\texttt{ltsfit} \citep{Cappellari2013a}, 
}

\facilities{Gaia, Sloan, WISE, ARC:3.5-m (DIS), GTC:10.4-m (HiPERCAM), Liverpool:2-m (IO:O), INT:2.5-m (WFC), FLWIO:2MASS, HST (STIS), GALEX, IRSA}

\bibliography{bib,references}
\bibliographystyle{aasjournal}

\end{document}